\documentclass{kapproc} 
\newcommand{\beq}{\begin{equation}}
\newcommand{\eeq}{\end{equation}}

\RequirePackage{graphicx}%
\RequirePackage{epsf}%
\input{psfig.sty}

\upperandlowercase
\let\footnote\savefootnote
\let\footnotetext\savefootnotetext 
\let\footnoterule\savefootnoterule 
\kluwerbib 

\begin{document}


\articletitle{A minimum hypothesis explanation for an IMF with a lognormal body and 
power law tail}


\chaptitlerunninghead{Power law tail in IMF}



\author{Shantanu Basu and C. E. Jones}
\affil{Department of Physics and Astronomy, The University of Western Ontario,\\ London, Ontario N6A 3K7, Canada}
\email{basu@astro.uwo.ca, cjones@astro.uwo.ca}





\begin{abstract}
We present a minimum hypothesis model for an IMF that resembles a lognormal distribution
at low masses but has a distinct power-law tail.
Even if the central limit theorem ensures a lognormal distribution
of condensation masses at birth, a power-law tail in the distribution arises due to 
accretion from the ambient cloud, coupled with a non-uniform (exponential) distribution
of accretion times.
\end{abstract}

\section{A Model for the Initial Mass Function}

\begin{figure}[ht]
\centerline{\psfig{file=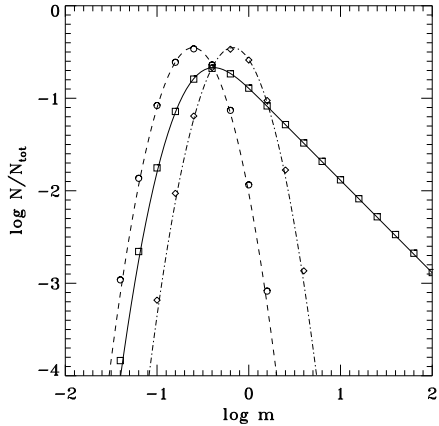,height=2.25in}}
\caption{Evolution of a distribution of masses $m$ undergoing accretion growth
$dm/dt = \gamma m$. Dashed line and circles: an initial lognormal probability
density function (pdf), with  
$\mu_0 = -1.40,\sigma_0 = 0.52$. Dash-dotted line and diamonds: pdf after
accretion growth of all masses for a time $t = \gamma^{-1}$.  
Solid line and squares: pdf if accretion lifetimes have 
a pdf $f(t) = \delta e^{-\delta t}$ and $\delta = \gamma$.
From Basu \& Jones (2004).}
\end{figure}

Observations of the field star IMF have long established the existence
of a power-law tail in the intermediate and high mass regime
(Salpeter 1955). 
Recent observations of stars within young embedded clusters (e.g., Muench
et al. 2002), have also established the existence of a low mass 
peak in the stellar mass distribution.
Submillimeter observations of dense protostellar condensations (e.g., Motte, Andr\'e, \&
Neri 1998) also imply a power-law tail in the intermediate and high mass regime.
Given the evidence for a peaked distribution, a natural explanation is to invoke the 
central limit theorem of statistics and argue that the IMF should be characterized by 
the lognormal probability density function (pdf)
\beq
\label{lognorm}
f(m) = \frac{1}{\sqrt{2 \pi} \sigma m} \exp \left[ - \frac{(\ln m - \mu)^2}
{2 \sigma^2} \right] 
\eeq
for the masses $m$, where $\mu$ and $\sigma^2$ are the mean and variance
of $\ln m$.
Now, assume that condensation masses are initially drawn from the above distribution
with mean $\mu_0$ and variance $\sigma_0^2$. Furthermore, if they accrete mass at a rate
$dm/dt = \gamma m$, (this may be a reasonable assumption at the 
condensation stage, but not for a protostar accreting from its parent core), and the 
accretion time $t$ has an exponential distribution with a pdf $f(t) = \delta
e^{-\delta t}$, the pdf of final masses is
\begin{eqnarray}
\label{newdistfcn}
f(m)  & =  & \frac{\alpha}{2} \exp \left[ \alpha \mu_0 + \alpha^2\sigma_0^2/2 \right] \: m^{-1 - \alpha}  \nonumber \\
& & \times \: {\rm erfc} \left[ \frac{1}{\sqrt{2}} \left( \alpha \sigma_0 - \frac{\ln m -\mu_0}{\sigma_0} \right) \right].
\end{eqnarray}
In this equation, $\alpha = \delta/\gamma$ is the dimensionless ratio of 
`death' rate to `growth' 
rate of condensations, and erfc is the complementary error function.
This analytically derivable three-parameter formula has the advantage of 
being near lognormal
at low masses but having an asymptotic dependence $f(m) \propto m^{-1-\alpha}$.
If the `death' and `growth' rates are both controlled by the parent cloud, we might
expect $\alpha \approx 1$, so that the distribution is Salpeter-like.
See Basu \& Jones (2004) for details, and references to other areas (e.g., distribution
of incomes, city sizes, and Internet file sizes) where similar ideas are applicable.



\end{document}